\documentclass[12pt]{article}
\usepackage{graphicx}
\usepackage{amsmath,amssymb}

  \def\be{\begin{equation}}
\def\ee{\end{equation}}  
\def\ba{\begin{array}{c}}

\def\ea{\end{array}} 
\def\bea{\begin{eqnarray}}
\def\eea{\end{eqnarray}}
\begin{document}
\title{Periodic square-well potential and spontaneous breakdown of
$PT$-symmetry}
%


\maketitle
\begin{center}{V\'i{}t Jakubsk\'y\footnote{jakub@ujf.cas.cz},
 Miloslav Znojil\footnote{znojil@ujf.cas.cz}}\\
{\'Ustav jadern\' e fyziky AV\v CR, 250 68 \v Re\v z, Czech
Republic }
\end{center}
\begin{abstract}
A particle moving on a circle in a purely imaginary one-step
potential is studied in both the exact and broken $PT$-symmetric
regime.
\end{abstract}

\section{Introduction}     

In textbooks on quantum mechanics one finds a lot of solvable
models. They mostly offer just a rough approximation to a physical
situation. At the same time, their simplicity enables us to avoid
some inessential technical difficulties. In this sense they
provide a basic insight in physical phenomena appearing in
complicated realistic systems.

Also the basic properties of $PT$-symmetric quantum mechanics
\cite{bender0} may be tested by the most elementary quantum
mechanical models. One of them has been proposed in~\cite{znojil1}
as a description of a particle moving in a purely imaginary
antisymmetric potential. In its time-independent Schroedinger
equation
 \be H\psi=\left[-\frac{d^2}{d
 x^2}+iZ\frac{x}{|x|}\right]\psi(x)=E\psi(x),\ \ x\in(-1,0)\cup(0,1)
  \label{squarewell}\ee
the Dirichlet boundary conditions were introduced at $x = \pm 1$.
The role of the growing non-Hermiticity $Z$ was studied and shown
to induce a spontaneous breakdown of the $PT$-symmetry at $Z
\approx 4.475$ (cf. also~\cite{znojil2}).

An interesting application of the latter model has been found
in~\cite{bagchi} where the whole supersymmetric hierarchy of
solvable potentials has been assigned to the most trivial
``zeroth-term" member (\ref{squarewell}) of the family.

In the present note we intend to replace the Dirichlet boundary
conditions at $x = \pm 1$ (mimicking a simple confined motion) by
their periodic alternative which would represent the slightly more
sophisticated motion of the particle along a circle.

Our main motivation stems from an observation~\cite{znojil3} that
a weakening of the Hermiticity (or, in our present language, of
$T-$symmetry \cite{BBJ}) to $PT-$symmetry may cause serious
difficulties in some exactly solvable models. In~\cite{znojil3}
this problem has been revealed during a study of angular
Schr\"{o}dinger equations with certain potentials of a
multiple-well shape over the circle. Unfortunately, even after a
replacement of these potentials by their schematic square-well
forms, the solution of the related bound-state problem proved more
or less purely numerical.

We see the main source of the latter difficulty in an
over-complicated structure of the underlying trigonometric secular
equations in the ``realistic" multi-square-well cases. For this
reason we intend to return to the ``non-realistic" single-step
potential in equation~(\ref{squarewell}) subject to the periodic
boundary conditions. We shall show and see that a graphical
analysis of such a problem remains tractable non-numerically.

\section{$PT$-symmetric regime}
Assuming that the energies are real, the solution
of~(\ref{squarewell}) may be sought in the form
 \bea\psi_1(x)&=&A_1e^{kx}+A_2e^{-kx},\
 \ \ \ \ \ \
 \ \ \ \ \ \ \ \ \ \ \ \
 x\in(0,1),\nonumber\\
  \psi_2(x)&=&B_1e^{k^*(x+1)}+B_2e^{-k^*(x+1)},\
  \ \ \ \ x\in(-1,0),
 \label{ansatz}\eea
where $k^2=-E+iZ$. Ambiguity in coefficients $A_i,\ B_i$ will be
eliminated by application of the following periodic boundary
conditions
 \be\psi_1(1)=\psi_2(-1),\ \ \psi'_1(1)=\psi'_2(-1),\ \
  \psi_1(0)=\psi_2(0),
 \ \ \psi'_1(0)=\psi'_2(0).\label{per.bound.cond.}\ee
Substituting~(\ref{ansatz}) into~(\ref{per.bound.cond.}), we
obtain a system of linear equations for unknown coefficients
$A_i,\ B_i$. It has non-trivial solution if and only if the
determinant of its matrix $W$ vanishes,
 \be \det
 W=4e^{2k}(-1+e^{2k})^2k^2=0\label{det}\ee
Dividing $k$ into its real and imaginary part and using the
following relations
 $$ E=s^2-t^2,\ 2st=Z,\ k=t+is,$$
we can rewrite~(\ref{det}) as
 \be
\det
 W'=
 4e^{-2t}(-1+e^{2t})^2t^2+\frac{2Z^2}{t^2}
 \left(-1+\cos\left(\frac{Z}{t}\right)\right)=0.\label{trepr}
  \ee
For large $t$ the first term in~(\ref{trepr}) is dominant and
exponentially grows to infinity, see Fig.1. The roots $t$ have a
positive upper bound and the energy is bounded from below,
consequently.

In the vicinity of $t=0$ the determinant in~(\ref{trepr}) exhibits
oscillations caused by the dominance of the non-positive second
term. Still, a positive perturbation caused by the first term
implies the existence of the infinitely many real and
non-degenerate nodal doublets in each oscillation at the
sufficiently small $t$. Empirically, this feature has been
observed in~\cite{znojil3} but its essence lied hidden in the
complicated form of the determinant.

\begin{figure}\begin{center}
\rotatebox{270}{\includegraphics*[210pt,256pt][413pt,539pt]{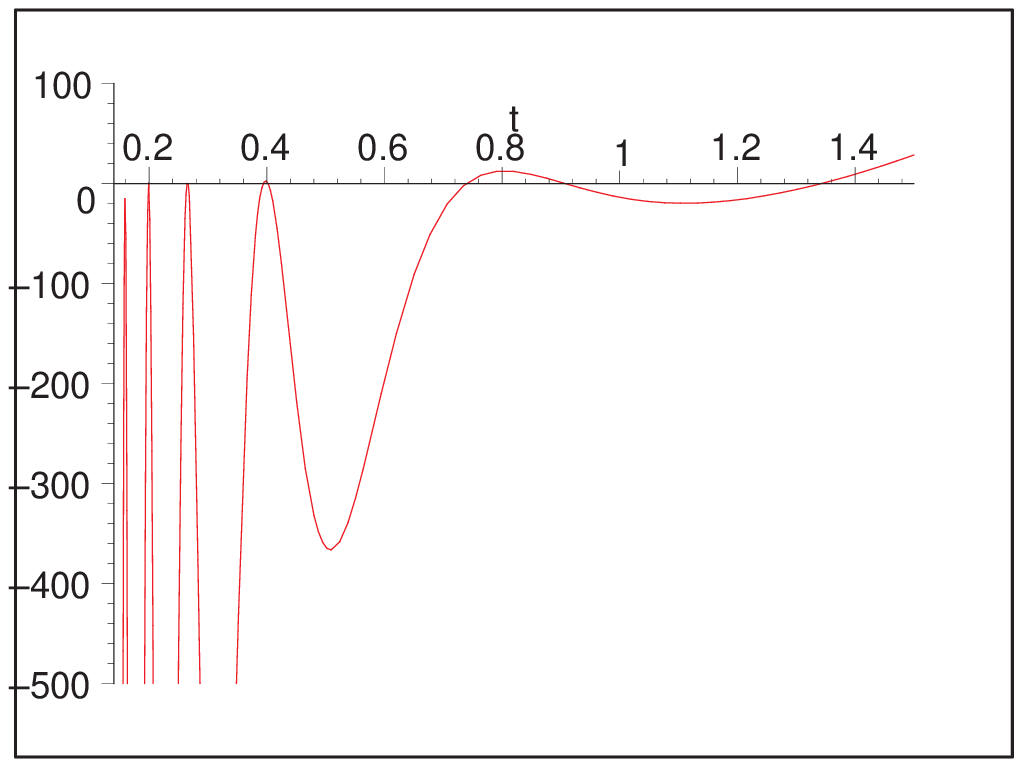}}
\caption{Determinant~(\ref{det}) in "t-representation",
$Z=5$.}\end{center}\end{figure}

To study the spectrum for infinitesimally small $Z\ (=2st)$, it is
convenient to rewrite~(\ref{det}) in its alternative
$s$-representation
 \be
  8s^2(-1+\cos(2s))+e^{-{\frac{Z}{s}}}\left(
 -1+e^{\frac{Z}{s}}\right)^2\frac{Z^2}{s^2}=0\label{srepr}
 \ee
Obviously, in the hermitian limit $Z\rightarrow 0$ the energies
coincide with the spectrum of the circular oscillator,
$E_n=s^2=\pi^2n^2$. We can ask how these ``unperturbed" energies
will be effected by a very small perturbation $Z>0$. It can be
expected that there will appear correction terms in energy
description. The secular equation~(\ref{det}) can be rewritten as
 \be (t\sinh t+ s\sin s)(t\sinh t-s\sin s)=0.\label{tsdet}\ee
We expect the correction to the hermitian case in the following
form
 \be s=n\pi+\rho(t),\ \mbox{where}\ \rho=\sum^{\infty}_0A_it^i.\label{skorekce}\ee
Substituting the ansatz into~(\ref{tsdet}) and comparing
coefficients at the corresponding powers of $t$, we get
 \bea\rho_{\pm} &=&\pm{\frac {(-1)^n}{n\pi }}t^2+ \left( -{\frac {1}{{n}^{3}{\pi }^{3}}}\pm{
 \frac {(-1)^n}{n\pi  }} \right) {t}^{4}\nonumber\\
 && + (-1)^n\left( \pm{\frac
 {2}{{n}^{5}{ \pi }^{5}}}-{\frac {(-1)^n}{3{n}^{3}{\pi
 }^{3}}}\pm{\frac {1}{6 n^3{\pi }^3}}\pm \frac{1}{120 n\pi} \right) {t}^{6}
 +..\ .\label{rho}\eea
In contrast to unperturbed spectrum of infinite square well, the
energies of $PT$-symmetric square-well are divided into two
families. They are
 \be E_n^+=(n\pi+\rho_+)^2-t^2,\ \ \ E_n^-=(n\pi+\rho_-)^2-t^2. \label{ptsymmE}\ee

\section{Violation of $PT$-symmetry}
 It has been observed in~\cite{znojil2} that as the
coupling $Z$ rises over a critical value $Z^{(crit)}$, two lowest
energy levels of the infinite square-well coalesce and become
complex conjugate simultaneously. This happens repeatedly as the
coupling rises, so that there exists a sequence of critical values
 \be Z_{0}^{(crit)}<Z_{1}^{(crit)}<Z_{2}^{(crit)}..<Z_{\nu}^{(crit)}
 \label{critseq}\ee
for which the corresponding energy pair $\{E_{2\nu},\
E_{2\nu+1}\}$ merges and becomes complex.

We can observe the very same situation in the case of periodic
boundary conditions. In Fig.2, an intersection of $Z=const$ with
the boarders of black and white area determines the root
of~(\ref{trepr}). Merging of the highest roots for rising coupling
$Z$ is then quite transparent.

\begin{figure}\begin{center}
\rotatebox{0}{\includegraphics*[115pt,358pt][294pt,554pt]{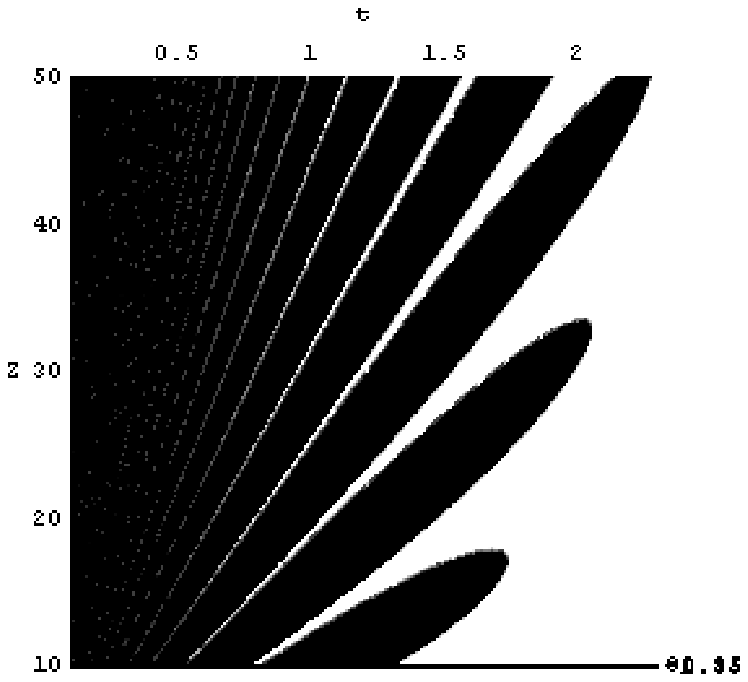}}
\caption{Determinant~(\ref{trepr}) vanishes on the border curve of
black and white area.} \end{center}\end{figure}

In order to study the system in the broken $PT$-symmetry regime,
we make the following ansatz of the wave function associated with
energy $E^{+}=E+i\epsilon$
 $$\psi_1(x)=A_1\sinh k(1-x)+A_2\cosh k(1-x),\ \ x\in(0,1)$$
 \be\psi_2(x)=B_1\sinh l^*(1+x)+B_2\cosh l^*(1+x),\ \
 x\in(-1,0)\label{ansatzbroken}\ee
where $k^2=-E+i\epsilon-iZ$ and $l^2=-E-i\epsilon-iZ$. In analogy
with exact $PT$-symmetry case, we substitute~(\ref{ansatzbroken})
into the boundary conditions and get a system of linear equations.
The corresponding secular equation
 \be 2k(1-\cosh k \cosh l^* )-\frac{k^2+{l^*}^2}{l^*}\sinh k \sinh l^*
  =0
 \label{detbroken}\ee
is complex valued and contains two complex parameters that are
mutually related
 $$k=s-it,\ l=p-iq \Rightarrow E=t^2-s^2=p^2-q^2,\
 \epsilon=pq-st.$$
It is convenient to make further re-parametrization
 $$s=K\sinh\alpha,\ t=K\cosh\alpha,\ p=K\sinh\beta,\ q=K\cosh\beta $$
Imaginary part of the energy is
 $$\epsilon=\frac{K^2}{2}(\sinh2\beta-\sinh2\alpha) $$
where
 $K=\sqrt{\frac{2Z}{\sinh2\alpha+\sinh2\beta}}. $
The parameters $\alpha$ and $\beta$ are solution
of~(\ref{detbroken}). Their values obtained numerically for
several fixed $Z$ can be found in Tab.1.

 \begin{table}[t]                          
\caption{Dependence of $E$ on the coupling $Z$ in the vicinity of
the
    first two critical values~$Z_{0}^{(crit)},\ Z_{1}^{(crit)}$.
    The first values of interaction correspond to preserved
    $PT$-symmetry so
    that parameters $\alpha$ and $\beta$ coincide. As the
    interaction grove over the critical value,
    the parameters diverse.}
\vspace{2mm} \small
\begin{center}                           
\begin{tabular}{|c|c|c|c|}           
\hline
                    $Z$&$\alpha$&$\beta$&$ReE$\\\hline
                    5.542309&0.474944&0.474944&5.041586\\\hline
                    5.542310&0.474653&0.474870&5.044077\\\hline
                    5.54232&0.474125&0.475399&5.044078\\\hline
                    5.54240&0.472878&0.476652&5.044080\\\hline
                    5.55&0.457619&0.492438&5.044371\\\hline
                    6&0.358129&0.622216&5.062183\\\hline
                    6.5&0.318347&0.693565&5.083353\\\hline\hline
                    17.90123&0.325829&0.325829&25.61820\\\hline
                    17.90124&0.325757&0.326139&25.60761\\\hline
                    17.90126&0.325540&0.326356&25.60762\\\hline
                    17.90200&0.323724&0.328189&25.60769\\\hline
                    17.95&0.308679&0.344308&25.61228\\\hline
                    19&0.253831&0.422062&25.71469\\\hline

\end{tabular}                            
\vspace{-1mm}
\end{center}                             
\end{table}                             

Comparing $\alpha$ and $\beta$, we can estimate the critical
values of interaction quite precisely. The first five values are
 \bea Z_{0}^{(crit)}\in
 (5.542309,5.542310),&&
 Z_1^{(crit)}\in(17.90123,17.90124)\nonumber\\
 Z_{2}^{(crit)}\in
 (33.54495,33.54495),&&
 Z_3^{(crit)}\in(51.20617,51.20618)\nonumber\\
 Z_4^{(crit)}\in(70.3093,70.3095).\eea
We can compare these results with infinite square-well.
In~\cite{znojil2}, the first two members of the
sequence~(\ref{critseq}) were determined as $Z_{0}^{(crit)}\in
(4.4748,4.4754)$, $ Z_1^{(crit)}\in(12.80154,12.80156)$. In our
case of periodic boundary conditions, the critical values of the
coupling seem to be risen. We propose that periodic boundary
conditions strengthen $PT$-symmetry of the system.

\section{Discussion and Outlook}
The paper was intended as a connection
between~\cite{znojil1},~\cite{znojil2} and~\cite{znojil3}. To meet
this intention, we studied solutions of~(\ref{squarewell}) with
periodic boundary conditions.

On one hand, the simpler choice of the potential allowed a deeper
insight into spectral behavior of periodic square-well, which was
the missing link in~\cite{znojil3}. We made a basic analytical
observation and found approximation of energies for very small
couplings $Z$. On the other hand, we could compare our results
with the ones corresponding to the infinite
square-well~\cite{znojil1},~\cite{znojil2}. This was interesting
mainly in the regime of broken $PT$-symmetry. After the
comparison, one concludes that $PT$-symmetry is weakened by
Dirichlet boundary conditions or vice versa, it is strengthened in
the circular domain.

There is a lot of opened questions. One can ask how is the energy
dependence of the critical interaction values. The similar task
has been solved in~\cite{bender1} for quartic harmonic oscillator
$H=-p^2+x^4+Aix$. It was shown that the relation
$a=|A|E^{-\frac{3}{4}}$ holds asymptotically for a certain
constant $a$.

Similarly to~\cite{znojil3},~(\ref{squarewell}) can be seen as an
angular Schroedinger equation of more dimensional problem. The
presented results could be also understood as a preliminary step
to more-dimensional models.

\bigskip
{\small Work supported from the budget of the AS CR project AV 0Z
1048901. Participation of M. Z. partially supported by grant GA AS
{C}R, grant Nr. 104 8302 }


\end{document}